\documentclass{andp2012}
\usepackage[english]{babel}
\usepackage{enumerate}
\usepackage{amsmath}
\title{Quantum Emulation of Extreme Non-equilibrium Phenomena with Trapped Atoms}
\author[S.\,V.\, Rajagopal]{Shankari V. Rajagopal\inst{1}}
\author[K.\,M.\,Fujiwara]{Kurt M. Fujiwara\inst{1}}
\author[R.\,Senaratne]{Ruwan Senaratne\inst{1}}
\author[K.\,Singh]{Kevin Singh\inst{1}}
\author[Z.\,A.\,Geiger]{Zachary A. Geiger\inst{1}}
\author[D.\,M.\, Weld]{David M. Weld\inst{1,}\footnote{Corresponding author\quad E-mail:~\textsf{weld@ucsb.edu}}}
\address[1]{Department of Physics and California Institute for Quantum Emulation, University of California, Santa Barbara, California 93106, USA}
\shortauthors{S.\,V.\,Rajagopal \emph{et al}}
\begin{abstract}
Ultracold atomic physics experiments offer a nearly ideal context for the investigation of quantum systems far from equilibrium.  We describe three related emerging directions of research into extreme non-equilibrium phenomena in atom traps: quantum emulation of ultrafast atom-light interactions, coherent phasonic spectroscopy in tunable quasicrystals, and realization of Floquet matter in strongly-driven lattice systems.  We show that all three should enable quantum emulation in parameter regimes inaccessible in solid-state experiments, facilitating a complementary approach to open problems in non-equilibrium condensed matter.
\end{abstract}
\shortabstract
\begin{document}
\maketitle

\section{Introduction}

Ultracold atoms in optical traps are increasingly being used as tools for the experimental study of many-body quantum systems important in other areas of physics~\cite{zwerger-review, lewenstein-review, blochdalibardQSreview}.  Such experiments, often called ``quantum emulation'' or ``quantum simulation,''  are made possible by the exquisite level of quantum control that can be exerted over cold-atom Hamiltonians.  
A major emerging area of interest in this field is the direct experimental investigation of non-equilibrium dynamics in tunable quantum systems~\cite{saenz-attosecondsimulatorPRA,holthaus-strongfieldsim,sengstock_moleculesimulator,polkovnikov-noneqRMP}.  Although much has already been done in this area, unexplored frontiers remain. In this article we define and discuss a part of this new frontier of quantum simulation experiments: the study of extreme non-equilibrium phenomena.

Since many if not most cold atom experiments have some non-equilibrium component, it is worth specifically  defining the scope of this work.  In the context of quantum emulation of condensed-matter physics, what we mean by an ``extreme'' non-equilibrium phenomenon is one which is difficult or impossible to realize in the solid state, for practical or fundamental reasons. Several new directions along these lines are possible.  For concreteness, we will focus on three such new directions for extreme non-equilibrium quantum emulation:
\begin{enumerate}
\item Quantum emulation of ultrafast processes.
\item Phasonic spectroscopy in tunable quasicrystals.
\item Floquet matter in strongly-driven lattices.
\end{enumerate}
Each of these directions concerns extreme non-equilibrium phenomena in the sense that cold atom quantum emulation techniques can move beyond what is possible in condensed-matter experiments.  In the sections below we define and discuss these new directions and present the results of initial calculations elucidating the advances that are within the reach of current experimental technology.  

\section{Quantum emulation of ultrafast atom-light interactions}
\label{ultrafastsec}
The first experimental direction we discuss is the use of trapped atoms to understand how electrons in atoms and solids respond to ultrafast strong electric fields.  Following earlier proposals~\cite{saenz-attosecondsimulatorPRA,holthaus-strongfieldsim}, we examine the possibility of quantitatively measuring the strong-field impulse response of artificial atoms and solids with variable pulse parameters, interactions, and lattice geometry, to advance understanding of the behavior of matter in pulsed-laser fields.  Such cold-atom quantum emulation of strong-field solid-state phenomena would probe some of the fastest processes in atomic physics using some of the slowest.  This  approach, which is complementary to existing experimental techniques, has the potential to address important open questions in strong-field physics and to investigate unexplored regimes of ultrafast-equivalent dynamics.  In this section, we will begin by motivating the proposed approach, proceed to a specific discussion of experimentally feasible ultrafast quantum emulation, and then present and discuss numerical models of simple initial experiments.

\subsection{Scientific motivation: ultrafast phenomena}

\begin{figure*}\begin{center}
\sidecaption
\includegraphics[width=.65\textwidth]{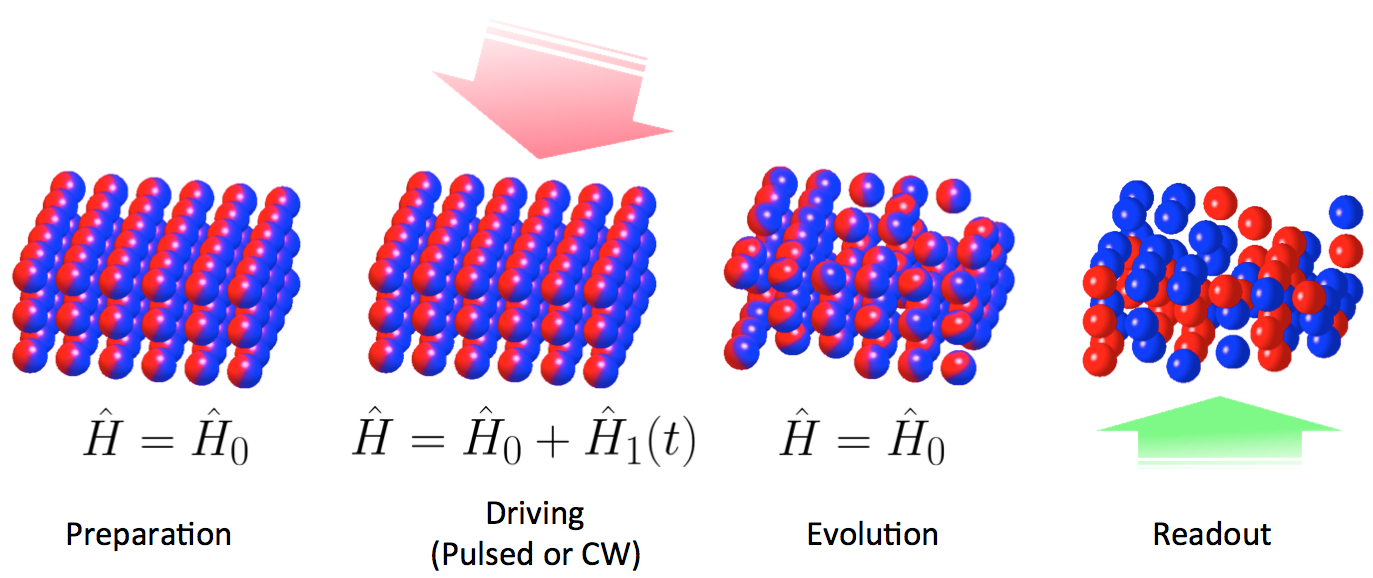}
\caption[QS]{\col General diagram of an experiment performing quantum emulation of ultrafast phenomena.  An initial state (e.g. a two-component Mott insulator) is prepared using standard techniques.  This system is then subjected to a strong pulsed or continuous drive, using optical or magnetic forces.  Subsequent evolution under the original Hamiltonian is followed by readout using absorptive, dispersive, or diffractive imaging.  }
\label{QGCdiagram}
\end{center}\end{figure*}

Revolutionary scientific and technological advances in ultrafast lasers have made strong-field ultrafast physics a vibrant and growing area of research~\cite{krausz-attosecond-review,corkumkrauszreview}.  Some exciting prospects include real-time imaging of valence electron motion, sub-optical-cycle control of solid-state conductivity, petahertz information processing, imaging of charge transfer in biomolecules, and production of Floquet-Bloch states of matter. However, especially in the solid state, our theoretical understanding of non-equilibrium strong-field physics is limited. Important open questions include the precise mechanism and timing of tunneling ionization, the effects of interactions and correlation, and the nature of the quantum-classical crossover. Accurate numerical treatments are infeasible even for moderate-sized single atoms, let alone solids. Simplified models of ultrafast dynamics are widely used \mbox{(\cite{corkum-model}, e.g.)}, but their range of applicability is unknown and validation is difficult. In addition to these theoretical challenges, physical limitations of pulsed laser experiments (on time resolution, pulse shape, maximum field intensity, net applied force, and readout) restrict the experimental investigation of some of the most exciting prospects in ultrafast physics: many processes occurring inside atoms and solids are still too fast for us to see or control.   Many of these limitations can be addressed or circumvented by the complementary technique of cold atom quantum emulation.

Quantum emulation experiments often rely on the analogy between neutral atoms in optical lattices and electrons in crystals. Although these two systems have vastly different energy densities, they can in many cases be described by equivalent Hamiltonians, which give rise to equivalent physics. This analogy has been fruitfully used to explore equilibrium solid-state phenomena from Mott insulators to antiferromagnets~\cite[e.g.]{greinerSFMI,greiner-AFspinchains}, and dynamical phenomena from quantum quenches to dynamical suppression of tunneling to integrability~\cite[e.g.]{greiner-revival, arimondo-shakenlattice,weiss-newtonscradle,hofferberth-coherencedynamics-short}. In the non-equilibrium context, the difference in energy scales leads to a rescaling of time: the processes underlying ultrafast phenomena like tunnel ionization occur over milliseconds rather than attoseconds, and are thus in a sense observable in ultra-slow-motion~\cite{saenz-attosecondsimulatorPRA}.  This greatly simplifies experimental access to the dynamics.  It is worth noting that the low energy density of cold atom experiments also enables freezing out of all internal degrees of freedom of the constituent atoms: preparation of an atomic sample in a single hyperfine state (or two, to model spin-1/2 particles) is a standard technique.  This level of quantum control of internal states is what allows for the emulation of a relatively structureless electron by a much-more-complex quantum object such as an alkali or alkaline-earth atom.

Despite the large temporal magnification factor and the ease of applying arbitrary driving fields to cold atoms with optical or magnetic field gradients, very little of the growing body of quantum emulation research has directly addressed ultrafast phenomena in atoms and solids.  Earlier work on accelerated optical lattices (e.g. \cite{raizen-wannierstark,raizen-tunneling}) can be understood as pioneering this approach.  A few trailblazing theoretical proposals have suggested the use of cold atoms to simulate gas-phase attosecond dynamics~\cite{saenz-attosecondsimulatorPRA} and multiphoton resonances in solids~\cite{holthaus-strongfieldsim}, and very recent experimental work has demonstrated the use of cold atoms to study femtosecond dynamics in an artificial benzene molecule~\cite{sengstock_moleculesimulator}. 

 A general diagram of an experiment implementing quantum emulation of ultrafast phenomena in a lattice appears in Fig.~\ref{QGCdiagram}.  Cold trapped atoms moving and interacting in an optical lattice are often approximately modeled with a modified Bose-Hubbard Hamiltonian:
\begin{equation}
\mathcal{H}_\mathrm{latt} = -J\sum_{\langle i j \rangle} (\hat{a}^\dag_i \hat{a}_j + \hat{a}^\dag_j \hat{a}_i) + \frac{U}{2}\sum_i \hat{a}^\dag_i\hat{a}^\dag_i \hat{a}_i\hat{a}_i + \sum_i \mu_i \hat{a}^\dag_i \hat{a}_i. \label{hammy}
\end{equation}
Here $J$ is the tunneling matrix element, $U$ is the onsite atom-atom interaction energy, $\mu_i$ is the site-dependent chemical potential resulting from the trap, $\hat{a}$ and $\hat{a}^\dag$ are ladder operators, and $\langle i j \rangle$ represents the sum over all nearest-neighbor pairs.  This model, representing a ``tunable artificial solid,'' supports rich physics including a well-studied quantum phase transition between superfluid and Mott insulator~\cite{greinerSFMI}.  Generalizations of this model, including to the case of multiple bands, Fermi statistics, or long-range or spin-dependent interactions, are experimentally possible.  External forces take the form of a space- and time-dependent chemical potential $\mu_i(t)$; these are the forces  used to perform quantum emulation of ultrafast processes.

While optical lattices enable access to complex and potentially useful ultrafast solid-state dynamics~\cite{holthaus-strongfieldsim}, many ultrafast experiments use individual atoms in the gas phase rather than solid materials.  Quantum emulation of single-atom phenomena can be done using a tight optical dipole trap, which plays the role of the Coulomb potential of the atomic nucleus~\cite{saenz-attosecondsimulatorPRA}.  In this context, the relevant Hamiltonian is
\begin{equation}
\mathcal{H}_\mathrm{trap} =  \sum_{i=1}^N \frac{\mathbf{p}_i^2}{2m} + \sum_{i=1}^N V_\mathrm{trap}(\mathbf{r}_i) + \mathcal{H}_\mathrm{int} + \sum_{i=1}^N V_\mathrm{applied}(\mathbf{r}_i,t),
\label{strongham}
\end{equation}
where $\mathbf{p}_i$ and $\mathbf{r}_i$ are the position and momentum of the $i^\mathrm{th}$ atom, $V_\mathrm{trap}$ is the trapping potential which emulates the Coulombic nuclear potential, $\mathcal{H}_\mathrm{int}$ describes atom-atom interactions, and $V_\mathrm{applied}$ is the time-dependent applied potential which emulates the effect of the electric field of an ultrafast laser. 

We emphasize that the goal of ultrafast quantum emulation experiments is not exclusively the preparation of systems with eigenstates and response functions mapping exactly to those of a particular atom, molecule, or solid (though interesting experiments along these lines are already being pursued, for example in the work on artifical benzene presented in Ref.~\cite{sengstock_moleculesimulator}).  A goal of equal or even greater importance is to advance our understanding of \emph{general} phenomena which are relevant to a broad range of far-from-equilibrium quantum systems.  We mention several examples of such general phenomena in the next subsection.  Especially with this goal in mind, the cold-atom context offers advantages complementary to those of ultrafast experiments.

Two main features of such quantum emulation experiments will allow them to explore new regimes of extreme non-equilibrium matter. The first unique feature is the ten-trillion-fold temporal magnification factor resulting from operation in the ultra-low-energy regime. The second feature, enabled by the well-stocked toolbox of ultracold atomic physics, is near-complete spatiotemporal control of the energy landscape.  Together, these features may enable a new approach to studies of ultrafast solid-state dynamics. The ability to study dynamics on ultrafast-equivalent timescales  has the potential to facilitate fully quantum-mechanical emulation of light-matter interactions in regimes well beyond the limits of existing theories and experiments. 
     
\begin{figure*}\begin{center}
\sidecaption
\includegraphics[width=.67\textwidth]{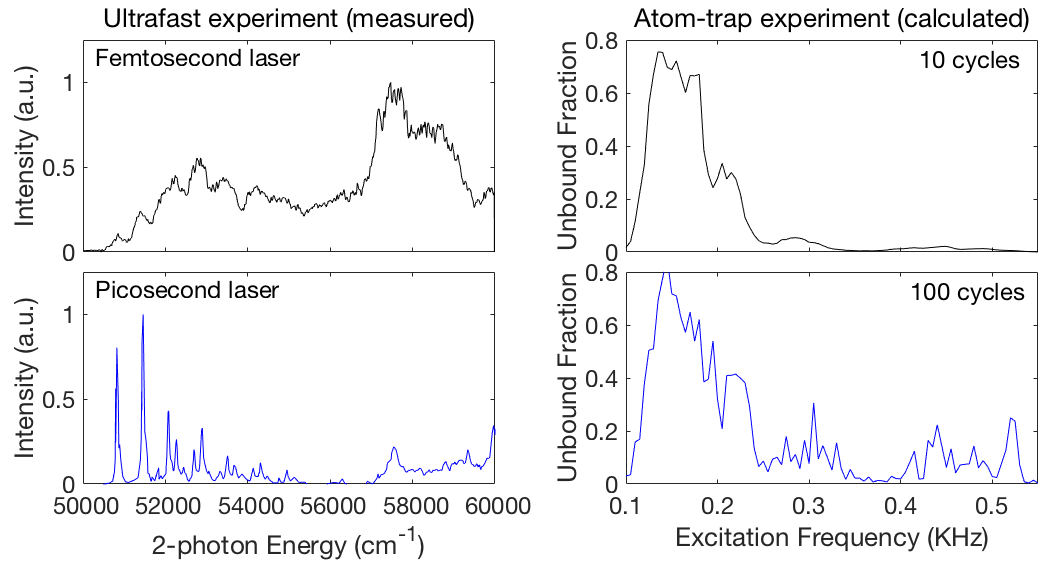}
\caption{\col Analogy between ultrafast and atom-trap experiments.  \textbf{Left:} Measured resonance-enhanced multiphoton ionization 
spectra of pyrazine-H4 at two different pulse times (data from Ref.~\cite{photoelectron_spectroscopy_fs_vs_ps}). \textbf{Right:} Calculated results from a strong-field quantum emulator consisting of $^{84}$Sr in a shaken optical dipole trap (waist $\sigma$ is 5 $\mu$m, depth $D$ is 100 nK, shaking amplitude $A_\mathrm{mod}$ is 1064 nm). The unbound fraction (which corresponds to the ionization yield) is plotted as a function of excitation frequency for two different pulse times.}
\label{SF1}
\end{center}\end{figure*}
 
 \subsection{Ultrafast quantum emulation experiments}
 
The first step in an ultrafast quantum emulation experiment is state preparation, in which well-understood cold atom techniques create an initial state of interest, such as a superfluid or Mott insulator, or simply macroscopic occupation of the ground state of a single trap.  The atoms in this state serve as quantum-mechanical models of bound electrons. The next step is illumination, in which strong pulsed or continuous drive fields are applied to the sample.  These fields serve to emulate the electric field of a pulsed laser. They can either be the result of magnetic field gradients or optical field gradients, and can be strongly spin-dependent.  Pump-probe pulse architectures are also straightforwardly implemented, thanks to the easily accessible experimental timescales.  In the next step, the fields are turned off and the system is allowed to evolve under its original Hamiltonian.  Finally (possibly after an additional probe pulse), the state is read out.  The cold atom context enables a set of readout techniques distinct from those possible in solid-state experiments: methods like time-of-flight momentum-space imaging, bandmapping, optical Bragg diffraction, and tomography allow precise projection and measurement of sample parameters at any time during or after a pulse.  This opens a path towards quantum emulation of a holy grail of ultrafast science: fully general attosecond-pulse attosecond-probe spectroscopy of complex correlated systems.  

The first and simplest measurements such emulators will enable are studies of strong-field tunnel ionization. In these experiments, bandmapping or simple time-of-flight imaging will be used to determine ``ionization yield'' (unbinding of atoms from the optical potential) as a function of the parameters of a few-cycle pulse. Even in the gas phase this cannot be considered a solved problem: the dynamics underlying ultrafast impulse response cannot be modeled exactly for atoms with more than two electrons.  Quantum emulation of such processes thus has an important role to play in elucidating the underlying physics, a role complementary to that of conventional ultrafast experiments.  To give an idea of future possiblities, we suggest a few examples of this complementarity. The effects of nuclear motion can be cleanly separated from the ionization physics, and tunneling ionization studies in the presence of only s-wave contact interactions can serve as a direct test of approximate theories which make use of a zero-range core potential~\cite{fewcycleionization-zerorange}.  Measurement of still-bound excited states after a pulse may enable new probes of frustrated tunneling ionization~\cite{saenz-frustratedtunneling}. The high temporal magnification and well-understood microscopic dynamics of the quantum emulator may provide valuable input into the long-running debate regarding tunneling times in field ionization experiments~\cite{dudovich-tunnelingtime,attoclock-tunnelingtiming,krausz-tunnelingtime}, by measuring ionization yield and the momentum of unbound atoms as a function of time at sub-cycle timescales. Studies of strong-field atomic stabilization~\cite{stronfieldstabilization} are an intriguing possibility. Another experiment enabled by ultrafast quantum emulation is the study of nonperturbative multiphoton resonances in driven lattices.  While barely accessible with modern pulsed lasers, such resonances can be studied cleanly and directly in strongly driven optical lattices~\cite{holthaus-strongfieldsim}.  Other possible targets for quantum emulation of few-cycle pulsed fields include studies of polarization dependence and multifrequency fields, and the effects of defects or inhomogeneities.  Polarization and multicolor excitation are straightforwardly implemented using two  modulated optical or magnetic field gradients with variable relative phase, while defects and inhomogeneities can be introduced by additional optical potentials or atoms of different spins or species.

Finally, interactions are a critical ingredient for a quantum emulator of ultrafast solid-state phenomena.  Because several options exist for emulation of electron-electron interactions, an array of exciting experiments is feasible.  Experiments with multiple spin states or multiple species are a natural possibility. One exciting goal is quantum emulation of the physics of recollision which underlies high-harmonic generation (HHG) in solids~\cite{Ghimire-HHGsolids}.  Although HHG itself depends upon the charged nature of the mobile constituents of matter, the recollision dynamics which result in HHG can be probed precisely by time-of-flight measurements of scattering halos (as in Ref.~\cite{ketterle-scatteringhalos}).  Studying halo structure as a function of pulse parameters and time should enable recollision studies with strongly-interacting isotopes.  Long-range interactions are more difficult to realize than contact interactions, but not impossible.  For example, transitions from the metastable triplet state of strontium can be used to engineer tunable long-range interactions via the exchange of mid-IR photons~\cite{sr-longrange-ints}.  Quantum emulation of ultrafast laser-solid interactions in the presence of arbitrarily tunable long-range interactions would represent a truly new capability for science.

\subsection{Calculated performance of ultrafast quantum emulators}

To provide specific insight into the operation of cold atom quantum emulators of ultrafast phenomena, we have  numerically modeled two simple non-interacting experiments by integration of the time-dependent Schr\"odinger equation.  In these and all calculations presented in this paper, the time-evolution operator is computed by a finite-difference method with periodic boundary conditions using the midpoint Crank-Nicolson method.  Appropriate time steps are chosen adaptively to control numerical error. For periodic drives, long-time numerical integration is achieved by iterative application of the single-period time-evolution operator.

We present the results here mainly in order to elucidate the relationship between ultrafast experiments and cold-atom experiments, and the relevance of the latter for performing quantum emulation of the former.  A crucial point underlying the motivation for experiments of the type we discuss is that no ab-initio theoretical calculation is capable of fully modeling the long-time dynamics of an interacting many-body quantum system driven far from equilibrium.  Because of this, a close dialogue between theory and experiment is necessary to advance our understanding of any complex quantum phenomenon (as has happened, for example, in recent years with experiments and theories probing the quantum critical point of the superfluid-Mott insulator phase transition).  Thus, the interaction-free calculation results we present are intended as a starting point for fruitful theory-experiment dialogue rather than a complete model of an interacting physical system. We emphasize that cold-atom quantum emulation is a technique which can \emph{complement} ultrafast experiments rather than replace them.  For more calculations and discussion of expected performance of ultrafast quantum emulation experiments, see also refs.~\cite{saenz-attosecondsimulatorPRA}~and~\cite{holthaus-strongfieldsim}.

The first experiment we analyze concerns the spectroscopic response of a bound system to a single ultrafast pulse.  This can be realized by loading degenerate bosons into the ground state of a tight Gaussian-beam optical dipole trap and observing the response to position-modulation of the trap.  Along a direction $x$ transverse to the trapping beam, the potentials in the Hamiltonian of Eq.~\ref{strongham} are then given by
\begin{equation}
V(x,t) = V_\mathrm{trap} + V_\mathrm{applied} = D e^{-\left(x-x_\mathrm{mod}(t)\right)^2/2 \sigma^2}, 
\label{strongpot}
\end{equation}
where $D$ is the optical trap depth, $\sigma$ is its waist, and
\begin{equation} 
x_\mathrm{mod}= A_\mathrm{mod} \sin(\omega_\mathrm{mod} t).
\end{equation}
Here $A_\mathrm{mod}$ is the modulation amplitude and $f_\mathrm{mod}=\omega_\mathrm{mod}/2\pi$ is the frequency of the modulation. 

The resulting inertial forces play the role of an electric field in an ultrafast experiment. A more exact realization of an effective electric field could be accomplished without substantially increased experimental complexity by using an applied optical intensity gradient or magnetic field gradient to provide the time-varying force. This can be understood as a cold-atom model of an ultrafast resonance-enhanced multiphoton ionization spectrum like those presented in Ref.~\cite{photoelectron_spectroscopy_fs_vs_ps}.  Near the Fourier-limited regime, the pulse duration affects the form and sharpness of the resulting spectra; this is demonstrated for femtosecond and picosecond excitation of pyrazine-H4 in the left panels of Fig.~\ref{SF1} (data are taken from Ref.~\cite{photoelectron_spectroscopy_fs_vs_ps}).  We have modeled a roughly equivalent cold-atom experiment by integrating the time-dependent Schr\"odinger equation in the time-varying potential of Eq.~\ref{strongpot}.  For simplicity we have neglected interactions.  The right panels of Fig.~\ref{SF1} show the dependence of unbound fraction (which corresponds to ionization yield) on pulse frequency for two different pulse lengths.  The point here is not to make quantitative comparisons between a single atom trap and a pyrazine molecule, but to support the possibility of useful quantum emulation by showing that similar Hamiltonians give rise to similar dynamical phenomena.

\begin{figure}\begin{center}
\includegraphics[width=.95\columnwidth]{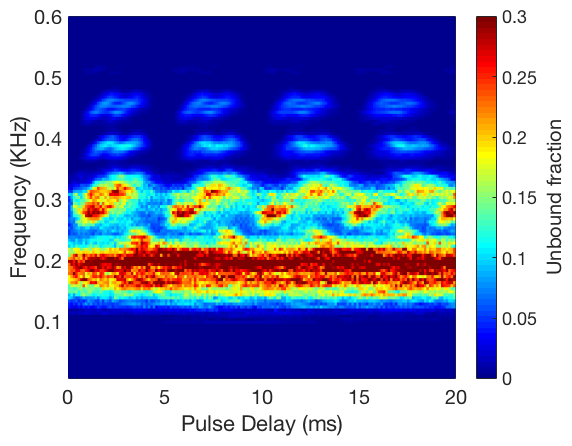}
\caption{Two-pulse ultrafast quantum emulation in an atom trap. Calculated unbound fraction (corresponding to ionization yield) after 2-pulse excitation is plotted versus pulse frequency $f_\mathrm{mod}$ and inter-pulse delay.  Note the interference effects in the dependence on pulse delay.  Trapped species is $^{84}$Sr, trap waist $\sigma$  is 5 $\mu$m, depth $D$ is 500 nK, and modulation amplitude $A_\mathrm{mod}$ is 2 $\mu$m for both 20-cycle pulses.}
\label{SF2}
\end{center}\end{figure}

The second experiment we have modeled is slightly more complex; it concerns the response of a bound system to a pair of ultrafast pulses separated by a variable delay time.  This is realized in the cold atom context in exactly the same way as the first experiment, but with a different modulation protocol consisting of two separated identical pulses.  Interference effects of varying the inter-pulse delay are clearly visible in the numerical results shown in Fig.~\ref{SF2}; even in a non-interacting system, the features of such spectra are not always simple to understand.  The two-pulse protocol could easily be extended to model phenomena involved in related ultrafast experiments such as those making use of the so-called RABBIT technique (reconstruction of attosecond beating by interference of two-photon transitions)~\cite{rabbit1,rabbit2}.  Further extensions involving varying polarization, multifrequency fields, partial-cycle pulses, and tunable interactions would be quite straightforward in the cold-atom context and would add substantially to the richness of the phenomena which could be investigated with this technique.  

\section{Phasonic spectroscopy}
The previous section focused on quantum emulation of phenomena which  occur when matter is illuminated by intense pulsed fields, and discussed the complementarity of conventional and emulator-based experiments.  The extreme tunability of cold-atom experiments also enables the realization of extreme nonequilibrium phenomena that cannot be studied in any other experimental context.  In this section we propose and discuss one such class of experiments: coherent phasonic spectroscopy of tunable quasicrystals.
 
 \subsection{Scientific motivation: quasicrystals and phasons} 
 
The formation, stability, excitation, and electronic structure of quasicrystals remain incompletely understood.  Open questions include the effects of electron-phason coupling, the nature of electronic conductivity or diffusivity, the spectral statistics,  the nature of strongly correlated magnetic states on a quasicrystalline lattice, topological properties of quasicrystals, and even the shape of the electronic wavefunctions~\cite{hofstadter-fibonacci-butterfly-2007,Thiel-dubois-QCcommentary,QCinAg,AFinQCs,ZilberbergQC,boundaryphenomena2,brouwerpaper,ames-QCs,hofstadter-superlattice-coldatom-proposal}. Just as phonon modes arise from discretely broken real-space translation symmetry, phason modes arise from broken translation symmetry in the higher-dimensional space from which all quasiperiodic lattices are projected~\cite{phasonreview,bakphasons,steinhardtphononsohasons}. Phasons have important but incompletely understood effects on thermal and electronic transport in real quasicrystals~\cite{phasonflips}. Because they involve long-range rearrangement of atoms, phasons are typically not dynamical degrees of freedom in solid-state quasicrystals; they are generally pinned to disorder or present as strain.  The influence of phasons is not understood in large part because of the experimental difficulty of disentangling the effects of domain walls, crystalline impurities, and disorder from those due to phason modes.  This is of interest not only for fundamental reasons, but also because of potential technological applications of quasicrystals' anomalous electrical and thermal transport characteristics.   

\begin{figure}\begin{center}
\includegraphics[width=.99 \columnwidth]{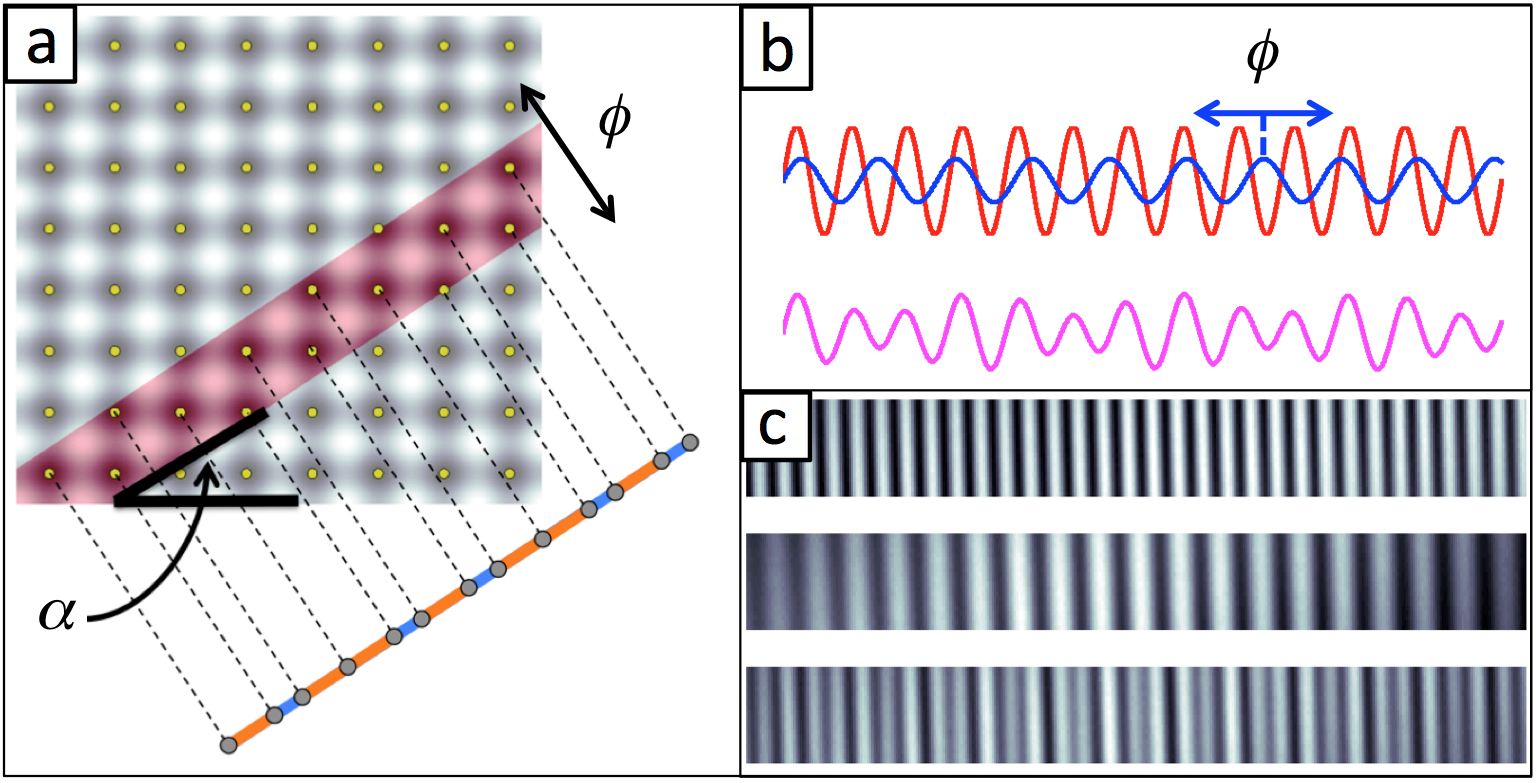}
\caption{\col Phasons in quasiperiodic optical traps.  \textbf{a:} Cut-and project construction of optical Fibonacci lattice, as described in Ref.~\cite{fibonacci-PRA}.  The cut angle $\alpha$ controls the quasiperiodicity, and translations in the direction labeled $\phi$ correspond to phasons.  \textbf{b:} Schematic of bichromatic optical lattice.  Here the phason parameter $\phi$ is the relative phase between the two single-frequency lattices. \textbf{c:} CCD images of tunable monochromatic and bichromatic lattices created with angled-beam interference.}
\label{phason1}
\end{center}\end{figure}
 
 \subsection{Coherent phason driving with trapped atoms}
The exquisite controllability of cold atoms makes them a natural choice for direct experimental investigation of such questions.  Quasiperiodic potentials that have been realized or proposed for cold atoms include  bichromatic lattices, well explored in the context of disorder-induced localization~\cite[e.g.]{inguscio-andersonloc,bloch-mblscience}, and the recently described generalized Fibonacci lattices~\cite{fibonacci-PRA}.  These two quasiperiodic structures are in fact topologically equivalent~\cite{equivalence}. The phason degree of freedom $\phi$ in Fibonacci lattices and bichromatic lattices is diagrammed in Fig.~\ref{phason1}.   

 A key experimental point is that both types of lattice enable driving of phasonic degrees of freedom via phase modulation of lattice beams.    This enables a simple procedure which is essentially impossible in a solid-state quasicrystal: measurement of the response of a  quasicrystal to driving phasonic modes at variable frequency and amplitude.  This would constitute a new form of lattice modulation spectroscopy, in which the modulation effectively occurs in the higher-dimensional space from which the quasiperiodic lattice is projected.  
  
In the simplest such experiment, energy absorption from the phasonic drive could be measured via standard time-of-flight calorimetry.  
 Better characterization of the effect of strong phason driving is possible with experiments that move beyond time-of-flight calorimetry and pursue measurements using Bragg spectroscopy and (in a deep lattice) doublon creation. This new kind of spectroscopy, impossible in other quasiperiodic systems, has the potential to allow clean and precise investigation of electron-phason coupling in quasicrystals.   
 
\subsection{Modeling phasonic spectra}

\begin{figure*}\begin{center}
\sidecaption
\includegraphics[width=0.7\textwidth]{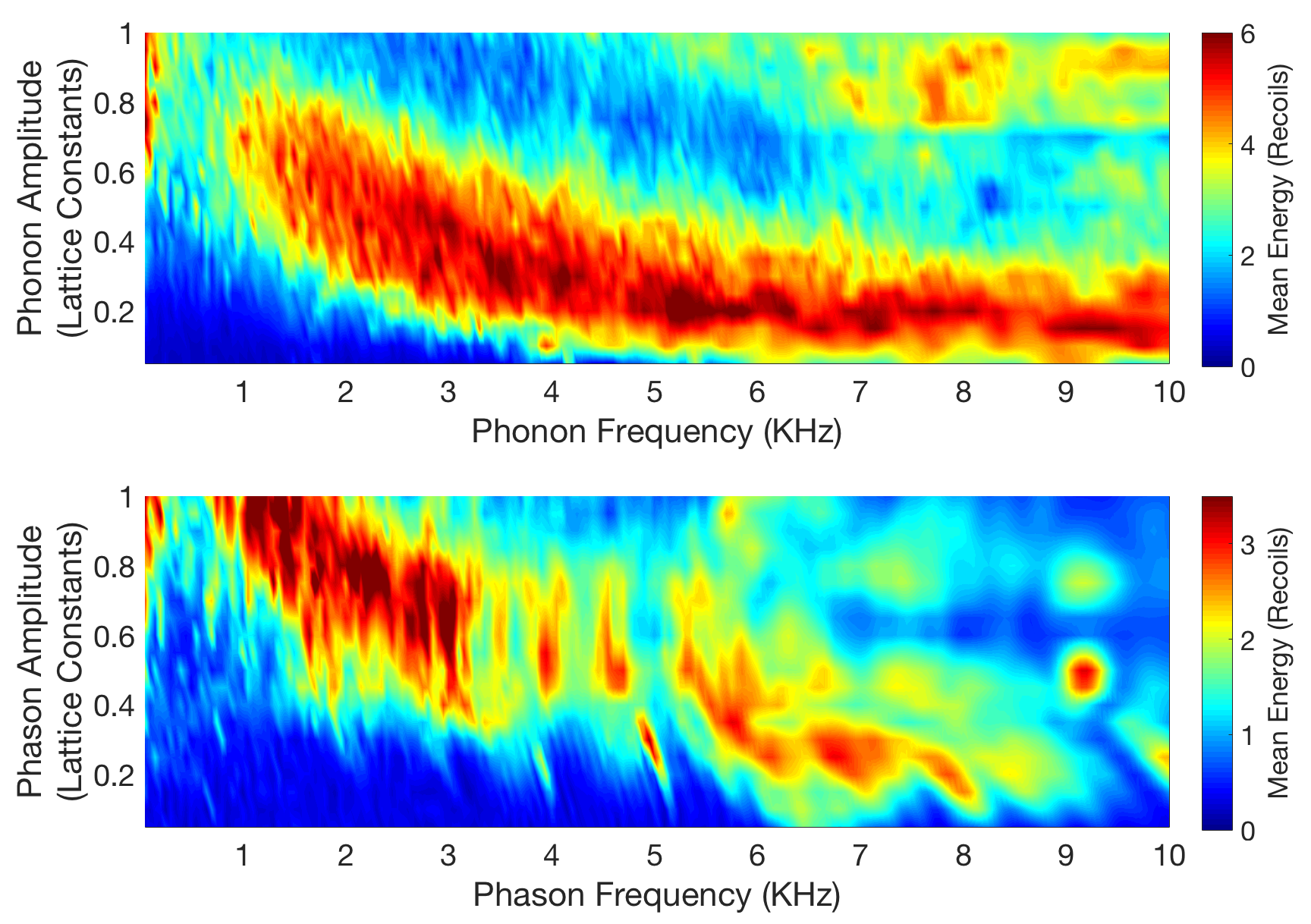}
\caption{ Coherent phasonic spectroscopy in quasiperiodic atom traps.  Plots show calculated energy absorption as a function of frequency $f$ and amplitude $A$ of phononic (top) and phasonic (bottom) excitation.  Calculations were performed for a 30-site bichromatic optical lattice with $k_S/k_L$ equal to the golden ratio $(1+\sqrt{5})/2$. Here $V_L = 1 E_R$, $V_S = 5 E_R$, and $k_S=2\pi/\lambda_\mathrm{YAG}$, with $\lambda_\mathrm{YAG}=1064\mathrm{nm}$, $E_R=\hbar^2 k_S^2/2m$, and $m$ taken to be the mass of $^{84}$Sr. Each  modulation was applied for 20 cycles, with the lattice phases varying in time as detailed in Eq.~\ref{phononeq} (top) and Eq.~\ref{phasoneq} (bottom).  All energies are referenced to the energy of the ground state.}
\label{phason2}
\end{center}\end{figure*}

Because phasonic spectroscopy is a fundamentally new tool, theoretical predictions for the results of such experiments are scarce.   In general, one expects significant differences from conventional ``phononic'' excitation as well as significant nonlinearities, particularly in the Fibonacci lattice, as large-amplitude phasonic displacements rearrange the lattice structure of the quasicrystal.  To demonstrate the potential of phasonic spectroscopy as a tool for investigating quasiperiodic quantum systems, we have calculated the response of a non-interacting cold-atom quasicrystal to phasonic and phononic excitation, again by integration of the time-dependent Schr\"odinger equation in a time-varying potential. The potential of a 1D bichromatic lattice can be written
\begin{equation}
V(x)=V_L \cos(k_Lx+\phi_L)+V_S \cos(k_Sx+\phi_S),
\end{equation}
where $V_L$, $k_L$, and $\phi_L$ ($V_S$, $k_S$ and $\phi_S$) are the amplitude, wavevector, and phase of the long (short) lattice.  Phasonic excitation of this lattice can be achieved for example by making only $\phi_L$ time-dependent:
\begin{equation}
\begin{aligned}
\phi_S(t) &= 0,\\
\phi_L(t) &= 2\pi A_\mathrm{phason} \sin(2\pi f_\mathrm{phason} t).
\end{aligned}
\label{phasoneq}
\end{equation}
Here we refer to $A_\mathrm{phason}$ and $f_\mathrm{phason}$ as the amplitude and frequency of the phasonic drive; these are the $y$ and $x$ axes of the bottom panel of Fig.~\ref{phason2}.  More conventional ``phononic'' driving can be achieved by translating the entire potential without changing the phase between the sublattices, for example by applying time-dependent phases as follows:
\begin{equation}
\begin{aligned}
\phi_S(t) &= 2\pi A_\mathrm{phonon} \frac{k_S}{k_L} \sin(2\pi f_\mathrm{phonon} t),\\
\phi_L(t) &= 2\pi A_\mathrm{phonon} \sin(2\pi f_\mathrm{phonon} t).
\end{aligned}
\label{phononeq}
\end{equation}
Here $A_\mathrm{phonon}$ and $f_\mathrm{phonon}$ are the amplitude and frequency of the phononic drive, and are the $y$ and $x$ axes of the top panel of Fig.~\ref{phason2}. The factor of $k_S/k_L$ is used so that all amplitudes are in units of the longer lattice period.

Fig.~\ref{phason2} shows the calculated mean energy in a bichromatic lattice with $k_S/k_L = (1+\sqrt{5})/2$ after both phononic excitation (following Eq.~\ref{phononeq}) and phasonic excitation (following Eq.~\ref{phasoneq}).  The phasonic and phononic spectra display numerous intriguing structures and are qualitatively different from one another.  Since our purpose is mainly to introduce the new technique of phasonic spectroscopy we will not explore them in detail here, but we emphasize that this unexplored technique is ripe for further investigation, both experimental and theoretical.

In keeping with our focus on extreme non-equilibrium phenomena, we have focused here on fast (diabatic) modulation of the phasonic degree of freedom.  It is worth noting that very slow phason modulation is also possible.  The resulting dynamics are of interest in part because adiabatic phasonic driving is expected to lead to long-range topological pumping of edge states~\cite{fibonacci-PRA,fibpumping}.  Both the nature of mass transport at the crossover between the diabatic and adiabatic regimes and the role of (Feshbach-tunable) interactions are further intriguing  frontiers for experimental investigation of the proposed extreme non-equilibrium technique of phasonic spectroscopy in finite-size systems.  

\section{Floquet matter in driven lattices} 

As our final example of an extreme non-equilibrium phenomenon accessible with trapped atoms, we describe a path to realization of a new state of matter in a strongly-driven optical lattice.  Such a realization would provide a clean, well-controlled experimental testbed for the study of Floquet states in driven many-body quantum systems, a topic of substantial current interest in a variety of contexts.

\subsection{Scientific motivation: Floquet matter}

A particularly exciting and rapidly-developing area of research in dynamical quantum systems is the study of non-equilibrium states of ``active quantum matter'' which arise only in the presence of strong driving~\cite{Nelson-THZcontrolofmatter}.   Recent advances have demonstrated that the phase structure which is the hallmark of equilibrium quantum matter can be generalized into the non-equilibrium regime~\cite{floquetmatter}. In the condensed-matter context, relevant recent experimental observations include non-equilibrium superconductivity in a BCS superconductor~\cite{noneqSCTHz}, a transient ferromagnetic state in an antiferromagnetic oxide~\cite{transientFM},  a metastable metallic phase in a thin-film dichalcogenide~\cite{metaastableinducedmetallicdichalcogenidestate}, Floquet-Bloch states on the surface of a topological insulator~\cite{gedki-floquetbloch}, and possible optically-enhanced superconductivity in high-$T_c$ cuprates~\cite{cavalleriYBCOemergent}.  A unified theoretical framework for understanding such phenomena does not yet exist. In these and other experiments, the details of the particular solids and driving fields used are crucially important, and disorder, decoherence, competing orders, and other complexities are unavoidably present.  Cold atom quantum emulation experiments are capable of cleanly focusing on only the most critical ingredients necessary for a many-body Floquet phase: a lattice, interactions, and strong driving~\cite{holthaus-floquetengineering,eckardt-drivenlatticereview}.  This makes cold atom quantum emulation complementary to solid-state experiments with pulsed lasers: it can open a new and uniquely clear window on complex emergent phenomena, and enable measurements which cannot be performed in any other way.

\subsection{Kapitza optical lattice}

We describe and discuss in this section a new state of matter which emerges in the presence of  sign-changing amplitude modulation of a lattice. Experimental realization of such a state would allow exploration of one of the simplest non-equilibrium Floquet phases accessible with cold atoms or indeed with any experimental system: a ``quantum Kapitza crystal.'' 

Ultrafast strong-field experiments generally make use of an electric field whose amplitude and possibly direction are a rapidly-varying function of time.  As argued in Section~\ref{ultrafastsec}, the toolbox of ultracold atomic physics allows emulation of such varying fields.  However, it also enables time-dependent tuning of  parameters like effective charge, which cannot in general be varied in solid-state experiments.   The nuclear charge which in solids is responsible for binding electrons to lattice sites appears in the cold atom context as an AC Stark shift due to an off-resonant optical trapping beam. Amplitude modulation of an optical lattice thus can be understood as an effective modulation of the nuclear charge in the solid-state analogy. Amplitude modulation is of course not a new capability; continuous and pulsed lattice modulation spectroscopy is a sufficiently standard technique that it is commonly used for calibration of optical lattice depths,  and  amplitude-modulated optical lattices have played a role in numerous pioneering experiments investigating, for example, quantum chaos~\cite{raizen-localization94,raizen-kickedrotor95,queensland_dynamicaltunneling,queensland-drivenpendulum,phillips-quantumresonances06,toronto-quantumresonanceskickedrotor}.  Control of ground-state equilibrium phase transitions using driving has already been predicted and demonstrated in cold atom experiments~\cite[e.g.]{eckardtholthaus-SFMIdriven,arimondo-coherentcontrol}, and modulated Hamiltonians are being explored as one avenue to strong synthetic gauge fields~\cite[and many references therein]{dalibard_drivensystems}.  

The behavior of quantum lattice systems exposed to \emph{sign-changing} amplitude modulation, however,  is not well investigated theoretically or experimentally. Far from being a simple extension of existing experiments, modulation intensities greater than 100\% actually open up a range of qualitatively new behavior.    As we argue below, continuous sign-changing  lattice modulation can stabilize the novel Kapitza state, which breaks translation symmetry in a fundamentally different way from any unmodulated phase. Experimentally, sign-changing modulation could for example be implemented by independent modulation of two orthogonally-polarized red-detuned lattices with a relative phase shift of $\pi$.  Cold-atom experiments in this regime could elucidate the dynamical emergence of the Kapitza state, and may give rise to other unexpected phenomena with solid-state relevance~\cite{cavalleriYBCOemergent}. 

\begin{figure}\begin{center}
\includegraphics[width=.88\columnwidth]{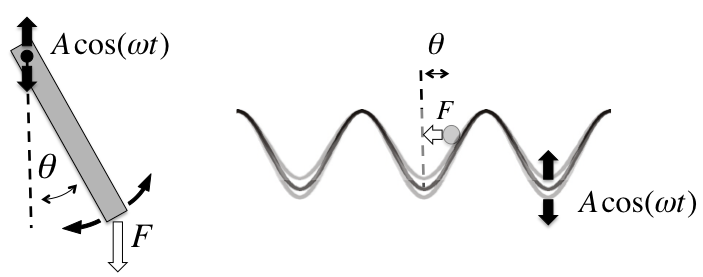}
\caption[Kapitzasetup]{\col Schematic of driven pendulum (left) and modulated optical lattice (right), highlighting equivalent variables. }
\label{kapitsasetup}
\end{center}\end{figure}

\begin{figure*}\begin{center}
\sidecaption
\includegraphics[width=.65\textwidth]{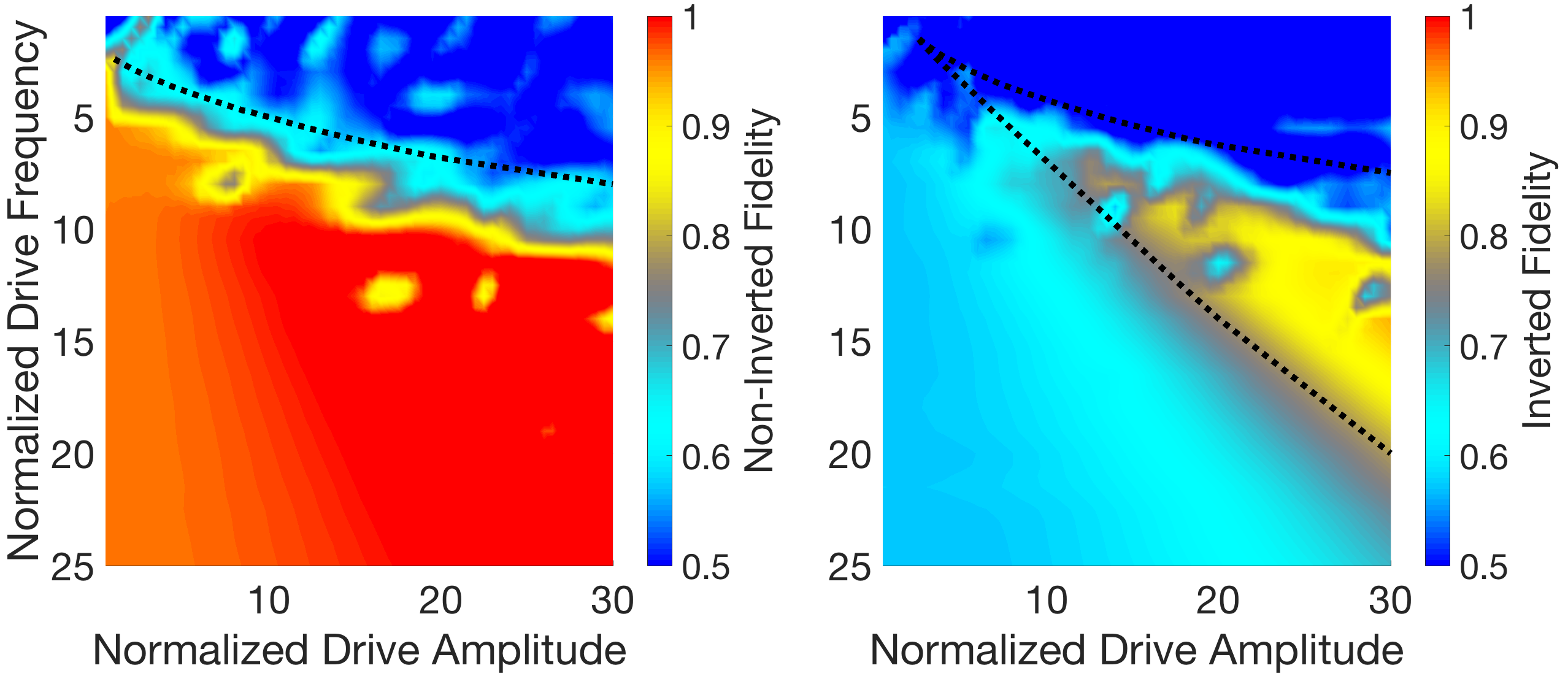}
\caption[kapitza]{\col  Regions of stability in a quantum Kapitza lattice, calculated by numerical TDSE integration. \textbf{Left:} Fidelity with which a non-inverted state is preserved, as a function of drive amplitude (in units of lattice depth) and drive frequency (in units of resonant frequency of the static lattice). \textbf{Right:} Fidelity with which an inverted state is preserved. In both panels, dashed lines indicate approximate boundaries of \emph{classical} stability regions.} 
\label{kapitza}
\end{center}\end{figure*}

Intuitively, the existence of the Kapitza state can be understood as a result of the fact that Hamiltonian for a classical particle moving in a 1D amplitude-modulated sinusoidal lattice is closely related to the Hamiltonian for a classical rigid pendulum in a gravitational field with an oscillating support (sometimes called a Kapitza  or ``inverted'' pendulum~\cite{kapitza}).  As diagrammed in Fig.~\ref{kapitsasetup}, the angular coordinate of the pendulum corresponds to the spatial coordinate of the particle.   Ignoring damping, the pendulum's classical equation of motion is 
\begin{equation}
\ddot{\theta}_p=-\left(\omega_0^2-\frac{C}{l}\omega^2\cos(\omega t)\right)\sin(\theta_p),
\label{pendulumeq}
\end{equation}
where $\omega_0=\sqrt{g/l}$ is the natural frequency, $m$ is the mass of the pendulum bob, $l$ is the length, $g$ is the acceleration due to gravity, $\theta_p$ is the angular displacement measured from the vertical down position, $C$ is the amplitude of the vertical pivot modulation, and $\omega/2\pi$ is the drive frequency of the pivot.  In the case of the amplitude-modulated lattice, the potential has the form
\begin{equation}
V\left(x,t\right)=-\cos\left(kx\right)\left(A+B\cos\left(\omega t\right)\right)\label{potideal},
\end{equation}
where $k$ is the lattice wavevector,  $A$ is the depth of the static lattice (which corresponds to the gravitational potential energy of the pendulum), $B$ is the strength of amplitude modulation of the time-varying part of the lattice, and $\omega$ is its frequency.  With the substitution $kx=\theta_l$, this potential gives rise to the classical equation of motion
\begin{equation}
\ddot{\theta}_l=-\left(\frac{k^2A}{m}+\frac{k^2B}{m}\cos\left(\omega t\right)\right)\sin(\theta_l),
\end{equation}
which is clearly of the same form as Eq.~\ref{pendulumeq}. Indeed, the classical equation of motion for both systems is a form of the Mathieu equation: 
\begin{equation} 
\ddot{\theta}+[\delta + \epsilon\cos(t)]\sin(\theta)=0,
\end{equation}
where $\delta$ is a reciprocal normalized modulation frequency and $\epsilon$ is a normalized modulation amplitude.  

The similarity between the equations of motion of the pendulum and the lattice allows both intuitive and quantitative understanding of the novel dynamical phenomena in the lattice.   At zero modulation amplitude, the rigid pendulum is of course stable in the downward-pointing configuration; this corresponds to the classical particle's stable location at the bottom of the time-averaged lattice potential.  It is, however, well known that for certain values of the driving amplitude and frequency, the Kapitza pendulum exhibits stabilization in the upward-pointing direction~\cite{kapitza}.  This corresponds to localization of the particle at the \emph{maximum} of the time-averaged lattice potential, and it only occurs for modulation amplitudes deeper than 100\%.  

Like the pendulum, the strongly-modulated lattice exhibits a rich classical stability diagram in which regions of stable amplitude-bounded motion (inverted, non-inverted, and bistable) are separated by regions of classically chaotic unbounded motion in which the particle absorbs energy from the driving field. The dashed lines in Fig.~\ref{kapitza} represent the classical boundaries between stable and unstable regions of the modulated lattice: the non-inverted state is stable below the dashed line in the left panel, and the inverted state is stable between the dashed lines in the right panel. Most intriguingly, a Kapitza crystal with half the original lattice constant (due to stable localization at both the maxima and minima of the time-averaged potential) emerges in the overlap of the inverted and non-inverted stability regions.  

\subsection{Modeling the quantum Kapitza state}

This treatment is so far entirely classical. As one might expect based on the correspondence principle, Kapitza stabilization is an extreme non-equilibrium phenomenon which can also occur in the quantum regime.  This has recently been discussed in a different experimental context in Ref.~\cite{demler-kapitza2}.  To elucidate the quantum-mechanical version of Kapitza stabilization in an amplitude-modulated lattice, we have calculated the response of degenerate lattice-trapped bosons to deep amplitude modulation, of the form given in Eq.~\ref{potideal}.  The initial wavefunction was chosen to be localized near either the time-averaged energy minima (``non-inverted'') or maxima (``inverted'').  Using  the numerical techniques described in section 2.3, the relevant Schr\"odinger equation was then integrated in the presence of strong amplitude modulation at a range of amplitudes and frequencies.  

The overlap of the beginning and ending atom distribution was used as a qualitative proxy for the stability of the chosen initial state. 
The results, shown in Fig.~\ref{kapitza}, support the classical intuition described above and indicate that the phenomenon of Kapitza localization persists in the quantum regime.  The regions of high ``fidelity'' to the initial state indicate a higher degree of stability of that state.  Dashed lines on the plots show the boundaries of \emph{classical} non-inverted and inverted stability regions; there is reasonably good overlap between quantum and classical calculations for these lattice parameters. Note in particular the overlap of the stablest areas in both panels of the figure--- in this part of the diagram, atoms are stable at both maxima and minima, resulting in a lattice constant reduced by a factor of 2 from the undriven case. Such a state would never be stable in the absence of driving; its observation would mark a milestone in the control of nonequilibrium states of quantum matter.

By smoothly changing the strength and frequency of the modulation, modulated-lattice experiments could move between regions where the atoms are stable at zero, one, or two locations in a single lattice period. Experiments on this new frontier could explore the non-equilibrium phase diagram experimentally, demonstrate stabilization of neutral atoms at potential energy maxima, create dynamical lattices with unprecedentedly close spacing of a quarter wavelength, and explore the effects of tunable interactions, dimensionality, dissipation, disorder, and tunneling on Kapitza localization.  

Various measurement protocols are possible.  The simplest concrete observable which can characterize Kapitza localization is energy absorption from the drive, measurable as a function of drive parameters via time-of-flight calorimetry.  Projection onto a static lattice and subsequent bandmapping could demonstrate ``inverted'' localization at potential maxima.  Optical Bragg diffraction~\cite{hulet-bragg,braggprl} could provide a sensitive potential probe of periodic ordering in a Kapitza crystal.  

In addition to  advances in our understanding of non-equilibrium phases of matter, possible applications of research into this extreme non-equilibrium phenomenon include switchable localization of atoms,  controlled wave-packet transport, and new tools for engineering nearest-neighbor interactions in a quantum simulator.  The effect of interactions on Floquet matter  is of particular interest,  is difficult to elucidate theoretically, and is potentially relevant to cutting-edge ultrafast solid-state experiments~\cite{cavalleriYBCOemergent}.  The addition of Feshbach-tuned interactions to the Kapitza lattice described above would enable an unprecedentedly well-controlled probe of interacting Floquet matter.  Strongly-modulated lattices thus offer a particularly simple path to realizing and exploring this extreme non-equilibrium phenomenon.

\section{Conclusion}
We have discussed and presented initial calculations elucidating three new directions for experimentally exploring extreme non-equilibrium phenomena using ultracold trapped atoms: quantum emulation of ultrafast strong-field physics, coherent phasonic spectroscopy of tunable optical-lattice quasicrystals, and realization of a new  state of matter in strongly-driven lattices.  All three directions seek to probe what one might call the ``non-equilibrium frontier'' by making use of unique features of trapped atoms such as extreme tunability and ability to be placed arbitrarily far from thermal equilibrium. All three are capable of exploring regimes of parameter space that cannot be straightforwardly attained in solid-state experiments.  Certainly other related directions of research in this area exist as well; we have focused on these three as an indication of the exciting open possibilities for cold-atom quantum emulation of extreme non-equilibrium phenomena.

\begin{acknowledgements}
The authors thank Alejandro Saenz, Sid Parameswaran, and Martin Holthaus for useful and interesting discussions, and Mikhail Lipatov and Toshi Shimasaki for a critical reading of the manuscript.  The authors acknowledge support from the National Science Foundation (CAREER award 1555313), the Army Research Office (award W911NF1410154), the Office of Naval Research (award N00014-14-1-0805), and President's Research Catalyst Award CA-15-327861 from the University of California Office of the President. 
\end{acknowledgements}


\bibliographystyle{andp2012}

\providecommand{\WileyBibTextsc}{}
\let\textsc\WileyBibTextsc
\providecommand{\othercit}{}
\providecommand{\jr}[1]{#1}
\providecommand{\etal}{~et~al.}

\end{document}